\documentclass[conference]{IEEEtran}
\IEEEoverridecommandlockouts
\usepackage{comment}
\usepackage{mathtools}
\usepackage{bbm}
\usepackage{graphicx}
\usepackage{lipsum}
\usepackage{rotating}
\usepackage{epsfig,graphics,amssymb}
\usepackage{amscd}
\usepackage{amsmath}
\usepackage{enumerate}
\usepackage{lipsum}
\usepackage{epstopdf}
\usepackage{mathtools}
\usepackage{theorem}
\usepackage{cite}
\usepackage{stfloats}
\usepackage{epsfig}
\usepackage{verbatim}
\usepackage{color}
\theoremstyle{plain} 
\theorembodyfont{\itshape}
 \theoremheaderfont{\scshape}
\newtheorem{theorem}{Theorem}
\newtheorem{lemma}{Lemma}
\newtheorem{remark}{Remark}

\newtheorem{definition}{Definition}
\newtheorem{proof}{Proof}
\newcommand{\new}{\color{black}}
\title{Expectation Propagation for Approximate Inference: \\Free Probability Framework}
\author{\IEEEauthorblockA{Burak \c{C}akmak and Manfred Opper \\Technical University of Berlin\\Berlin 10587, Germany \\Email:\{burak.cakmak,manfred.opper\}@tu-berlin.de
\thanks{Both authors are co-first authors.}	
}
}

\begin{document}
\def\mathlette#1#2{{\mathchoice{\mbox{#1$\displaystyle #2$}}%
                               {\mbox{#1$\textstyle #2$}}%
                               {\mbox{#1$\scriptstyle #2$}}%
                               {\mbox{#1$\scriptscriptstyle #2$}}}}
\newcommand{\matr}[1]{\mathlette{\boldmath}{#1}}
\newcommand{\RR}{\mathbb{R}}
\newcommand{\CC}{\mathbb{C}}
\newcommand{\NN}{\mathbb{N}}
\newcommand{\ZZ}{\mathbb{Z}}
\newcommand{\bfl}[1]{{\color{blue}#1}}
\newcommand{\sign}{\ensuremath{\mathrm{sign}}}
\maketitle
\begin{abstract}
We study asymptotic properties of expectation propagation (EP) -- a method
for approximate inference originally developed in the field of machine learning.
Applied to generalized linear models, EP iteratively computes a multivariate Gaussian approximation to the exact posterior distribution. 
The computational complexity of the repeated update of covariance matrices severely limits the application of EP to large problem sizes. In this study, we present a rigorous analysis by means of free probability theory that allows us to overcome this computational bottleneck if specific data matrices in the problem fulfill certain properties of asymptotic freeness. 
We demonstrate the relevance of our approach on the gene selection problem of a microarray dataset.
\end{abstract}
\section{Introduction}
We consider the problem of approximate (Bayesian) inference for a latent vector $\matr w$ with its posterior probability density function (pdf) of the form $p(\matr w \vert\matr y, \matr X)\propto p(\matr w) p(\matr y \vert \matr X \matr w)$ where $p(\matr w)$ is a prior pdf and $\matr X$ is an $N\times K$ matrix. Here all variables are real-valued. This probabilistic model often is referred to as ``generalized linear model''\cite{Nelder}. For a (multivariate) Gaussian $p(\matr w)$, it covers e.g. many Gaussian process inference problems in machine learning. But we are also interested in more general cases where both factors can be non-Gaussian. Such models naturally appear in signal recovery problems where a signal described by a vector $\matr w$ is linearly coupled $\matr z = \matr X\matr w$ and then passed through a channel with the output signal vector $\matr y$ and the channel pdf $p(\matr y\vert \matr z)$. 

Actually, it is rather convenient to introduce the latent vector $\matr X \matr w$ as an auxiliary vector $\matr z$ \cite{albert1993} and study the posterior pdf of the latent vector $\matr \theta\triangleq(\matr w, \matr z)$\footnote{By abuse of notation, for column vectors $\matr w$ and $\matr z$ we introduce the conjugate column vector notation by $(\matr w, \matr z)\triangleq(\matr w^\dagger, \matr z^\dagger)^\dagger$ with $(\cdot)^\dagger$ denoting the (conjugate, in general) transposition.}  which factorizes according to
\begin{equation}
p(\matr \theta\vert\matr y, \matr X) \propto \underbrace{p(\matr w)p(\matr y\vert \matr z)}_{p_1(\matr \theta)}\underbrace{\delta(\matr z-\matr X\matr w)}_{p_2(\matr \theta)}\label{joint}.
\end{equation}
Here,
for short we neglect the dependencies on $\matr y$ and $\matr X$ in the notations of $p_{1}(\matr \theta)$ and $p_{2}(\matr \theta)$, respectively.

We are interested in the method of (Gaussian) expectation propagation (EP) \cite{Minka1,OW5}.  It approximates the marginal posterior pdf $p(\theta_i\vert\matr y, \matr X)$ by $q(\theta_i),\forall i$ which is a marginal of the Gaussian pdf
\begin{equation}
q(\matr \theta)
\propto\underbrace{e^{-\frac{1}{2}\matr \theta^\dagger \matr \Lambda_1 \matr \theta +\matr \theta^\dagger \matr \gamma_1}}_{f_1(\matr \theta)}
\underbrace{e^{-\frac{1}{2}\matr \theta^\dagger \matr \Lambda_2 \matr \theta +\matr \theta^\dagger \matr \gamma_2}}_{f_2(\matr \theta)}\label{apdf}.
\end{equation}
Here, the parameters of $f_{1}(\matr \theta)$ and $f_{2}(\matr \theta)$ are obtained from the \emph{expectation consistency} (EC) conditions
\begin{equation}
\mathbb{E}[\{g(\matr \theta)\}]_{p_1\cdot f_2}=\mathbb{E}[\{g(\matr \theta)\}]_{f_1\cdot f_2}=\mathbb{E}[\{g(\matr \theta)\}]_{f_1\cdot p_2}\label{EC}
\end{equation}
with the expectation notation $\mathbb{E}[(\cdot)]_{m_1\cdot m_2}\triangleq\int (\cdot){p}(\theta){\rm d}\theta$ for a pdf $p(\theta)\propto m_1(\theta)m_2(\theta)$. The set $\{g(\matr \theta)\}$ is specified in such a way that $\mathbb{E}[\{g(\matr \theta)\}]_{f_1\cdot f_2}$ yields the set of all natural parameters of the Gaussian pdf $q(\matr \theta)$. For example,  the mean and the covariance are required in general, i.e. $\{g(\matr \theta)\}=\{\matr \theta, \matr \theta\matr \theta^\dagger\}$.  

The conventional approach in machine learning is to restrict the matrices $\matr \Lambda_1$ \& $\matr \Lambda_2$ to diagonals. Consequently, the EC conditions in \eqref{EC} are fulfilled  for   
\begin{equation}
\{g(\matr \theta)\}=\{\matr \theta, \matr \theta\odot \matr \theta\}~\text{ with }~(\matr \theta\odot \matr \theta)_i= (\matr\theta)_i^2.
\end{equation}
The \emph{diagonal-EP} approach has shown excellent performances for many relevant machine learning problems \cite{OW0,Minka1,Rasmussen}.

In the diagonal-EP case where $\matr \Lambda_1$ \& $\matr \Lambda_2$ being (arbitrary) diagonal, an iterative algorithm associated with the EP fixed-point equations \eqref{EC} will require a matrix inversion operation at each iteration. This leads to the iterative algorithm to have a cubic computational complexity, say $O(K^3)$, per iteration which makes a direct application of the method to large systems problematic. A simple way to resolve this problem is to consider the scalar-EP approach \cite{OW5} in which the diagonal matrices $\matr \Lambda_i$ for $i=\{1,2\}$ are restricted to $
{\small \matr \Lambda_i=\left(\begin{array}{cc}
	\lambda_{iw}{\bf I}_K & \matr 0 \\
	\matr 0 & \lambda_{iz}{\bf I}_N
	\end{array}   \right)}$ where ${\bf I}_K$ is $K\times K$ identity matrix. Consequently, the EC conditions in \eqref{EC} are fulfilled for
\begin{equation}
\{g(\matr \theta)\}=\{\matr w,\matr z, \matr w^\dagger \matr w, \matr z^\dagger \matr z\}.
\end{equation}
Such approach has recently been attracted considerable attention in the information theory community \cite{Kab14,Samp, Ma,SAMPI,Fletcher} because of its strong connection with the approximate message passing techniques \cite{Kabashima,Donoha,Rangan}. Moreover, important theoretical analyses on the trajectories of the scalar-EP fixed-point algorithms have been developed in the context of the linear observation model \cite{Keigo,rang,Burak17}.

The diagonal-EP method may stand as a plain approximation -- at first sight. But it actually coincides with the powerful cavity method introduced in statistical physics \cite{OW0,Adatap}. On the other hand, though the asymptotic consistency of the scalar-EP method with the powerful replica ansatz \cite{Tulino13} is now well-understood \cite{Keigo,rang,Burak17}, it is still unclear how to obtain the method per se from an explicit mathematical argumentation. 

In this study, we introduce a profound theoretical approach by means of free probability theory \cite{Hiai,mingo2017free} to relate the diagonal- and scalar- EP methods. Specifically, we show that if the data matrix $\matr X$ fulfills certain properties of asymptotic freeness both methods become asymptotically equivalent. Though the underlying asymptotic freeness conditions will be typically approximations in practice, we are nevertheless convinced that they can be reasonable approximations for many real-world problems. Indeed, in Section~5 we demonstrate the relevance of our free probability approximations on the gene selection problem of a microarray dataset. 	
	
\subsection{Related Works}\label{rworks}
The main technical step in the proof of the aforementioned contribution is to show the following: Let $\matr \Lambda$ and $\matr J$ be $n\times n$ real diagonal and Hermitian matrices, respectively, and $\matr Z$ be an $n\times n$ diagonal matrix and independent of $\matr \Lambda$ and $\matr J$. The diagonal entries of $\matr Z$ are independent and composed of $\pm 1$ with equal probabilities. If $\matr J$ and  $\{\matr \Lambda, \matr Z \}$ are asymptotically free we have $\small{[(\matr \Lambda+\matr J)^{-1}]_{ii}-(\Lambda_{ii}+v)^{-1}\to 0}, \forall i$ as $n\to \infty$  where $v$ is an appropriately computed scalar. Such a ``local law'' for addition of free matrices has been shown by \cite{Bao2017} (see also \cite{Kargin15}) but for $\matr J$ being invariant, i.e. $\matr J=\matr U \matr {\tilde \Lambda}\matr U^\dagger$ where $\matr U$ is Haar matrix that is independent of diagonal $\matr {\tilde \Lambda}$.  

While invariant matrices are important ensembles for the asymptotic freeness to hold, there are important matrix ensemble that are not invariant but the asymptotic freeness can still holds. For example, consider $\matr J=\matr U \matr {\tilde \Lambda}\matr U^\dagger$ as above. Then, by substituting the Haar matrix $\matr U$ with randomly permuted Fourier or Hadamard matrices the asymptotic freeness still holds, see \cite{Greg} for further information.

The proof technique by \cite{Bao2017} is based on the \emph{probabilistic analysis} of certain operations of Haar measure. Our proof technique is \emph{algebraic} and considerably simpler which could be a useful approach to obtain such \emph{local law} property of random matrices for different problems. Indeed, in our context we need to extend the above result to a more-involved matrix model in which our algebraic approach is convenient. On the other hand, the weak aspect of our method is that we prove the convergence {\new in the $L^2$ norm sense (see the explanation in the first paragraph of Section~IV)} while \cite{Bao2017} shows the convergence point-wise. This weakness is potentially remediable, but it exceeds the scope of the present contribution and is left for future study. 

Finally, in \cite{Burak16} we and the co-authors provide a non-rigorous (and rather complicated) approach to the problem.  
\subsection{Outline} 
This paper is organized as follows: Section 2 briefly presents the concept of ``asymptotic freeness'' (in the almost sure sense) in random matrix theory, see \cite{ralfc} for a detailed exposition. Section~3 presents a more explicit form of the fixed-point equations in \eqref{EC} for the diagonal- and scalar- EP methods. Our main result is presented in Section~4. Section~5 is devoted to the simulation results. Section~6 gives a summary and outlook. The proof of the main result is located in the Appendix. 
\section{Asymptotic Freeness}
In probability theory we say the real and bounded random variables $A$ and $B$ are independent if for all integers $n,m\geq 1$ we have $\mathbb{E}[(A^{n}-{\mathbb E}[A^{n}]) (B^{m}-{\mathbb E} [B^{m}])]=0$. This role of independence in expectations $\{\mathbb E[\cdot]\}$ of products of (commutative) random variables is analogous to the role of ``asymptotic freeness'' in asymptotic normalized traces $\{\phi(\cdot)\}$ of products of (non-commutative) matrices with $\phi(\matr A)\triangleq\lim_{N\to \infty}\frac{1}{N}{\rm tr}(\matr A)$ for an $N\times N$ matrix $\matr A$. Specifically, we say the matrices $\matr A$ and $\matr B$ are asymptotically free if for all $k\geq 1$ and for all integers $n_1,m_1,\cdots n_k,m_k\geq 1$ we have~\cite{Bao2017}
\begin{equation}
\phi \left( \prod_{i=1}^{k}(\matr A^{n_i}-\phi (\matr A^{n_i})\matr {\bf I} ) (\matr B^{m_i}-\phi (\matr B^{m_i})\matr {\bf I})\right)=0. \label{deffreeness}
\end{equation}
Note that the adjacent factors in the product must be polynomials belonging to different matrices. This gives the \emph{highest degree of non--commutativity}
in the definition of asymptotic freeness. We next present the concept of asymptotic freeness in a more general and formal setup\cite{ralfc}. To this end we define a non-commutative polynomial set in $p$ matrices of order $n$:
\begin{align}
\mathcal P_{n}(\matr A_1,\cdots,\matr A_p)\triangleq \left\{\sum_{i=1}^{\infty}\alpha_i \prod_{k=1}^n \matr A_{1}^{l_{i,k,1}}\cdots\matr A_{p}^{l_{i,k,p}}\nonumber \right. \\ \left. : \alpha_i \in \RR \text{ and } \left(\sum_{k=1}^{n}l_{i,k,j}\right)\in [1,n],~\forall i,j   \right\}.
\end{align}
For example $\mathcal P_{2}(\matr A, \matr B)$ consists of matrices of the form 
\begin{align*}
{\small \alpha_1 \matr A^2 \matr B^2+\alpha_2\matr A\matr B^2\matr A+\alpha_3\matr A\matr B\matr A\matr B+\dots+\alpha_{13}\matr A^2+}\\
{\small+\alpha_{14}\matr A\matr B+\alpha_{15}\matr B\matr A+\alpha_{16}\matr B^2+\alpha_{17}\matr A+\alpha_{18}\matr B+\alpha_{19}{\bf I}}.
\end{align*}
\begin{definition}\cite{Hiai}\label{deffree} The sets $\mathcal Q_1\triangleq\{\matr A_1,\ldots, \matr A_{\rm a}\},\mathcal Q_2\triangleq\{\matr B_1,\ldots, \matr B_{\rm b}\},\ldots, \mathcal Q_r$ form an asymptotically free family if
	\begin{equation}
	\phi \left(\prod_{i=1}^k\matr Q_i\right)=0\label{cond}
	\end{equation}
	whenever $\matr Q_{i}\in \mathcal P_{\infty}(\mathcal Q_{n(i)})$, $n(1)\neq n(2)\neq\cdots \neq n(k)$ with $\phi(\matr Q_{i})=0$ for all $i\in [1,k]$.
\end{definition}
We refer the reader to \cite[Section~4.3]{ralfc} for some important examples of asymptotically free matrices. 
\section{The EP Fixed-Point Equations}
In the sequel we present a more explicit form of the fixed-point equations in \eqref{EC} for the diagonal- and scalar- EP methods: Firstly, as regards the first expectation term in \eqref{EC} we define for convenience the vectors
\begin{align} 
\matr \eta \triangleq {\mathbb E}[\matr \theta]_{p_1\cdot f_2}~\text { and }~\matr \chi\triangleq {\mathbb E}[(\matr \theta-\matr\eta)\odot(\matr \theta-\matr\eta)]_{p_1\cdot f_2}.\label{eta}
\end{align}
Secondly, as regards the third expectation term in \eqref{EC} we define  
\begin{equation}
\matr \mu\triangleq\mathbb{E}[\matr w]_{f_1\cdot p_2}~\text { and }~\matr\Sigma\triangleq {\mathbb E}[(\matr w-\matr\mu)(\matr w-\matr\mu)^\dagger]_{f_1\cdot p_2}. \label{third}
\end{equation}
Here, we point out the following immediate remarks that $\mathbb{E}[\matr z]_{f_1\cdot p_2}=\matr X\matr \mu$ and ${\mathbb E}[(\matr z-\matr X\matr\mu)(\matr z-\matr X\matr\mu)^\dagger]_{f_1\cdot p_2}=\matr X\matr \Sigma \matr X^\dagger$. 
In particular, by invoking the standard Gaussian integral identities we have the explicit expressions of $\matr \Sigma$ and $\matr \mu$ as 
\begin{align}
\matr \Sigma&=({\matr \Lambda_{1w}}+\matr X^\dagger {\matr \Lambda_{1z}}\matr X)^{-1} \label{defa}\\
\matr \mu&=\matr \Sigma(\matr \gamma_{1w}+\matr X^\dagger \matr \gamma_{1z})\label{defb} 
\end{align}
where, for notational convenience, associated with the
vectors $\matr w$ and $\matr z$ we consider the EP parameters for $i=\{1,2\}$ as
\begin{equation}
\matr \gamma_{i}=(\matr \gamma_{iw},\matr \gamma_{iz})~~ \text{and} ~~ \matr \Lambda_{i}={\small\left(\begin{array}{cc}
	\matr \Lambda_{iw} & \matr 0 \\
	\matr 0 & \matr \Lambda_{iz}\end{array}\right)} ~\label{decom}
\end{equation}
with $\matr \gamma_{iw}\in \RR^{K\times 1}$ and $\matr \Lambda_{iw}\in \RR^{K\times K}$. Note that for the scalar-EP $\matr \Lambda_{iw}=\lambda_{iw}{\bf I}$ and $\matr \Lambda_{iz}=\lambda_{iz}{\bf I}$ where ${\bf I}$ is the identity matrix of appropriate dimension. Similar to \eqref{decom} we also write $\matr \eta=(\matr \eta_w,\matr \eta_z)$ and $\matr \chi=(\matr \chi_{w},\matr \chi_{z})$. Thus, the fixed-point equations \eqref{EC} for the diagonal- and scalar- EPs read of the form \cite{Burak16} 
\begin{align}
(\matr\eta_s)_i&=\frac{(\matr\gamma_{1s}+\matr\gamma_{2s})_i}{(\matr \Lambda_{1s}+\matr \Lambda_{2s})_{ii}}={\begin{cases}
	(\matr \mu)_{i} & \hspace{0.1cm}\qquad{s}={w} \\
	(\matr X \matr \mu)_{i} & \hspace{0.1cm} \qquad{s}={z}\end{cases}}
\label{fix1} \\
(\matr\chi_s)_{i}&=\frac{1}{(\matr \Lambda_{1s}+\matr \Lambda_{2s})_{ii}}={\begin{cases}
	(\matr \Sigma)_{ii} & ~{s}={w} \\
	(\matr X \matr \Sigma \matr X^\dagger)_{ii} & ~{s}={z}  \end{cases}} \label{fix2} \\
\langle \matr\chi_{s}\rangle&=~\frac{1}{~~\large\lambda_{1s}+\lambda_{2s}~}~={\begin{cases}
	{\rm Tr}(\matr \Sigma) & {s}={w}~~~~ \\
	{\rm Tr}(\matr X\matr \Sigma \matr X^\dagger)& {s}={\rm z}~~~~ \end{cases}} \label{fix3}.
\end{align} 
Here, for a $R\times R$ matrix $\matr A$ and a $R\times 1$ vector $\matr a$ we define
 ${\rm Tr}(\matr A)\triangleq \frac 1 R{\rm tr}(\matr A)$ and $\langle \matr a \rangle\triangleq \frac1 R\sum_{r=1}^{R}a_r$. Note that \eqref{fix1}~\&~\eqref{fix2} and \eqref{fix1}~\&~\eqref{fix3} constitute the fixed-point equations of the diagonal- and the scalar- EPs, respectively. 

\subsection{Cubic versus quadratic computational complexities}
An iterative algorithm to solve the diagonal-EP fixed-point equations, i.e. \eqref{fix1}~\&~\eqref{fix2}, has an $O(K^3)$ computational complexity per iteration (here we assume that $N/K=O(1)$) as it requires to update the matrix inverse \eqref{defa} at each iteration. 

On the other hand, an iterative algorithm for the scalar-EP fixed-point equations, i.e. \eqref{fix1} \& \eqref{fix3}, can have $O(K^2)$ computational complexity per iteration. Specifically, by computing the singular value decomposition of $\matr X$ before the iteration starts, the cubic computational complexity of updating \eqref{defa}, i.e. $({\lambda_{1w}}{\bf I}+{\lambda_{1z}}\matr X^\dagger\matr X)^{-1}$, at each iteration reduces to a quadratic computational complexity, see e.g. \cite{Schniter16}. 

Within the scalar-EP method it is also possible to circumvent the need for the singular value decomposition. In short, this can be done by bypassing the need for matrix inverse \eqref{defa} in fulfilling \eqref{fix1} (see next subsection) and fulfilling \eqref{fix3} by means of the so-called R-transform of free probability (see \cite{Burak16}) which is in general can be estimated from spectral moments ${\rm Tr}((\matr X^\dagger \matr X)^k)$ up to some order \cite{Jolanta}.
 
\subsection{An equivalent representation of \eqref{fix1}}\label{subTAP}
From \eqref{defb}, $\matr \eta_{w}=\matr \mu$ in \eqref{fix1} holds if, and only if 
\begin{equation}
\matr \gamma_{1w}=({\matr \Lambda_{1w}}+\matr X^\dagger {\matr \Lambda_{1z}}\matr X)\matr \eta_{w}-\matr X^\dagger \matr \gamma_{1z}.
\end{equation}
Combining this equality with the first equality of \eqref{fix1}, i.e. $(\matr \Lambda_1+\matr \Lambda_2)\matr\eta=(\matr\gamma_1+\matr\gamma_2)$, we have
\begin{equation}
\matr \gamma_{2w}=(\matr \Lambda_{2w}+\matr X^\dagger\matr \Lambda_{2z}\matr X)\matr \eta_{w}-\matr X^\dagger\matr \gamma_{2z}. \label{gam1}
\end{equation}
Here, we express $\matr \gamma_{2z}$ implicitly via $\matr\eta_{z}$: Consider the scalar functions $g_{i}((\matr\gamma_{2z})_i)\triangleq(\matr\eta_{z})_i$ $\forall i$, see \eqref{eta}. One can show that $g_{i}(x)$ is a strictly increasing function of $x$ and thereby its inverse (with respect to functional decomposition) $g_{i}^{-1}$ exists. Thus, $(\matr \gamma_{2z})_i=g_{i}^{-1}((\matr X\matr \eta_w)_i)$ holds if, and only if 
\begin{equation} 
\matr \eta_z=\matr X\matr \eta_w\label{gam2}. 
\end{equation}
In summary, \eqref{fix1} holds if, and only if \eqref{gam1} \& \eqref{gam2} hold. 

\section{The Main Result}
For $\matr \Lambda_{2}$ fixed since \eqref{gam1} \& \eqref{gam2} do not depend on $\matr \Lambda_{1}$ both the diagonal- and scalar- EP methods share the same fixed-point equations. The fixed-point equations of $\matr\Lambda_{2}$ differ however. That of diagonal-EP is given by \eqref{fix2} and that of scalar-EP is given by \eqref{fix3}. We next show under certain asymptotic freeness conditions that {\new $\{(\matr\Lambda_{2w}-\lambda_{2w}{\bf I})_{kk}:k\leq K \}\overset{L^2}{\rightarrow}0$ and $\{(\matr\Lambda_{2z}-\lambda_{2z}{\bf I})_{nn}:n\leq  N \}\overset{L^2}{\rightarrow}0$} where $\matr\Lambda_{2w}$ \&  $\matr\Lambda_{2z}$ are as in \eqref{fix2} and ${\lambda_{2w}}$ \& ${\lambda_{2z}}$ fulfill~\eqref{fix3}. {\new Here, for $\matr a\in \RR^n$ $\{a_1,\cdots,a_n\}\overset{L^2}{\rightarrow}A$ stands for $\lim_{n\to \infty} \mathbb E [(A_n-A)^2]=0$ where the distribution function of $A_n$ is defined as the empirical distribution function of $\{a_1,\cdots,a_n\}$\footnote{\new Specifically, the empirical distribution function of the entries of $\matr a\in \RR^n$ is given by ${\rm F}_{\matr a}(x)\triangleq\frac 1 n \vert \{a_i: a_i\leq x\} \vert$.}, i.e. $A_n$ converges in the $L^2$ norm to a random variable $A$.}
\begin{theorem}\label{main}
Let $\matr \Lambda_{1w}$ and $\matr \Lambda_{1z}$ be a $K\times K$ and an $N\times N$ real diagonal and invertible matrices, respectively, and $\matr X$ be an $N\times K$ matrix. Furthermore, let $\matr \Lambda_{2w}$ and $\matr\Lambda_{2z}$ be respectively a $K\times K$ and an $N\times N$ diagonal matrices whose diagonal entries are introduced through 
\begin{equation}
\frac{1}{(\matr \Lambda_{1s}+\matr \Lambda_{2s})_{ii}}={\begin{cases}
	(\matr \Sigma)_{ii} & ~{s}={w} \\
	(\matr X \matr \Sigma \matr X^\dagger)_{ii} & ~{s}={z}  \end{cases}}
\end{equation}
where $\matr \Sigma \triangleq({\matr \Lambda_{1w}}+{\small\matr X}^\dagger {\matr \Lambda_{1z}}\matr \small{\matr X})^{\small-1}$. 
Moreover, let $\matr Z_{w}$ and $\matr Z_{z}$ be a $K\times K$ and an $N\times N$ diagonal matrices, respectively, and independent of $\matr \Lambda_{1w}$, $\matr \Lambda_{1z}$ and $\matr X$. The diagonal entries of both $\matr Z_w$ and $\matr Z_z$ are independent and composed of $\pm 1$ with equal probabilities. In general, we assume that $\matr \Lambda_{1w}$, $\matr \Lambda_{1z}$ and $\matr X^\dagger \matr X$ have almost surely (a.s.) a limiting eigenvalue distribution (LED) each as $N,K\to \infty$ with $N/K=O(1)$.
\begin{enumerate}
\item If $\matr X^\dagger \matr \Lambda_{1z}\matr X$ has a LED a.s. and is asymptotically free of $\{\matr \Lambda_{1w},\matr Z_{w}\}$ a.s. {\new there exists a deterministic variable $\lambda_{2w}$ such that we have a.s.
	\begin{equation}
	\{(\matr\Lambda_{2w})_{11},\cdots,(\matr\Lambda_{2w})_{KK}\}\overset{L^2}{\rightarrow}\lambda_{2w} \label{result1}
	\end{equation}}whenever $(\matr \Lambda_{1w}+\lambda_{2w}{\bf I})^{-1}$ has a.s. an asymptotically bounded spectral norm. 
\item If $\matr X \matr \Lambda_{1w}^{-1}\matr X^\dagger$ has a LED a.s. and is asymptotically free of $\{\matr \Lambda_{1z},\matr Z_{z}\}$ a.s. {\new there exists a deterministic variable $\lambda_{2z}$ such that we have a.s. 
\begin{equation}
	\{(\matr\Lambda_{2z})_{11},\cdots,(\matr\Lambda_{2z})_{NN}\}\overset{L^2}{\rightarrow}\lambda_{2z}\label{result2}
	\end{equation}}whenever $(\matr \Lambda_{1z}+\lambda_{2z}{\bf I})^{-1}$ has a.s. an asymptotically bounded spectral norm.
\item If the pairs $\matr X\matr X^\dagger$~\&~$\matr \Lambda_{1z}$ and $\matr X^\dagger\matr X$~\&~$\matr \Lambda_{1w}$ form a.s. an asymptotically free family each then $\matr X^\dagger \matr \Lambda_{1z}\matr X$ and $\matr X \matr \Lambda_{1w}^{-1}\matr X^\dagger$ have a LED each a.s., respectively, and under the premises of 1) and 2) $\lambda_{2w}$ in \eqref{result1} and $\lambda_{2z}$ in \eqref{result2} are the solutions of the fixed-point equations 
	\begin{equation}
	\hspace{-0.3cm}{\small \chi_{s}=\frac{1}{\large\lambda_{1s}+\lambda_{2s}}={\begin{cases}
			{\phi}(({\lambda_{1w}}{\bf I}+\lambda_{1z}{\small\matr X}^\dagger \matr \small{\matr X})^{\small-1}) & {s}={w}\\
			{\phi}(\matr X({\lambda_{1w}}{\bf I}+\lambda_{1z}{\small\matr X}^\dagger \matr \small{\matr X})^{\small-1} \matr X^\dagger)&{s}={\rm z}\end{cases}}\label{solsoflambdas}} 
	\end{equation}
	where $\chi_{w}\triangleq\phi(\matr \Sigma)$ and $\chi_{z}\triangleq \phi(\matr X\matr \Sigma\matr X^\dagger)$.
\end{enumerate}
\proof{See Appendix.}
\end{theorem}

Theorem~\ref{main} is self-contained in the sense that its premises do not make reference to any of EP methods as $\matr \Lambda_{1w}$ and $\matr \Lambda_{1z}$ are some arbitrary diagonal matrices. Moreover, instead of assuming that e.g. $\matr X^\dagger \matr \Lambda_{1z}\matr X$ is asymptotically free of $\matr \Lambda_{1w}$ \emph{for any} $\matr \Lambda_{1w}$  it is sufficient to assume that $\matr X^\dagger \matr \Lambda_{1z}\matr X$ is asymptotically free of $\{\matr \Lambda_{1w},\matr Z_{w}\}$. 

The asymptotic freeness conditions in Theorem~\ref{main} hold if $\matr X$ is bi-invariant (i.e. the probability distribution of $\matr X$ is invariant under multiplications from left and right with any independent orthogonal matrices), $\matr \Lambda_{1w}$, $\matr \Lambda_{1z}$ and $\matr X$ are independent of each others and $\matr \Lambda_{1w}$, $\matr \Lambda_{1z}$ and $\matr X^\dagger \matr X$ have LEDs with uniformly bounded spectral norms \cite{Collins-a}. Of course, the condition of independence becomes artificial in an application of EP. Nevertheless, as the diagonal entries of $\matr \Lambda_{1w}$ and $\matr \Lambda_{1z}$ have a limiting (deterministic) distribution each we may treat the underlying matrices as asymptotically independent.

Note that \eqref{solsoflambdas} may not yield unique solutions for $\lambda_{2w}$ and $\lambda_{2z}$. Thus, for the asymptotic equivalence between the diagonal- and scalar- EP methods to hold we need to assume that the later yields unique solutions for $\lambda_{2w}$ and $\lambda_{2z}$. It is interesting to investigate a ``region of stability'' analysis \cite{AT} where the solutions offered by  \eqref{solsoflambdas} are unique. This could clarify the region of system parameters where both methods yields reliable solutions. But we consider this as out of the scope of the current contribution. 
\section{Simulation Results on Real Data}
To demonstrate that our free probability assumptions are applicable in practice, we address a novel application of EP to gene selection\footnote{The work \cite{LobatomAEP} addressed a factorizing EP method to the problem which neglects the dependencies of the latent variables in the posterior and yields a cruder approximation.}. This problem is based on microarray data \cite{lee_gene} which are used to analyze the tissue samples taken from $N$ patients. The data consists of $\{\matr X,\matr y\}$ where the matrix element $X_{nk}$ stands for the measurement of the expression level of the $k$th gene for the $n$th patient for $n\in [1,N]$ and $k\in [1,K]$, and the vector $\matr y=(y_1,\cdots, y_N)^{\dagger}$ contains binary labels where $y_n=-1$ or $y_n=1$ indicate whether the $n$th patient is healthy or has cancer. Given the data, the problem is to identify relevant genes for the cancer type of interest and to diagnose new patients. We model the inference problem~as~\cite{albert1993}
\begin{equation}
\matr y=\sign(\matr X\matr w+\matr \epsilon). \label{model}
\end{equation}
Here, $\matr \epsilon$ stands for a white Gaussian noise vector and $\matr w$ for the vector of regression parameters, where $w_k$ corresponds to the $k$th gene. The number of genes $K$ is typically in the range from $6000$ to $60000$. To model the prior assumption that only a small subset of genes are relevant for the cancer type of interest \cite{golub_99} we postulate
a \emph{sparsity} promoting prior distribution for $\matr w$ in the form of a spike-and-slab (aka Bernoulli-Gaussian) prior as $p(\matr w)=\prod_{k=1}^{K}(1-\rho)\delta(w_k)+\rho\mathcal N(w_k\vert 0,1)$ for $\rho\in (0,1]$.

\begin{figure*}[t]
	\includegraphics[width=\textwidth]{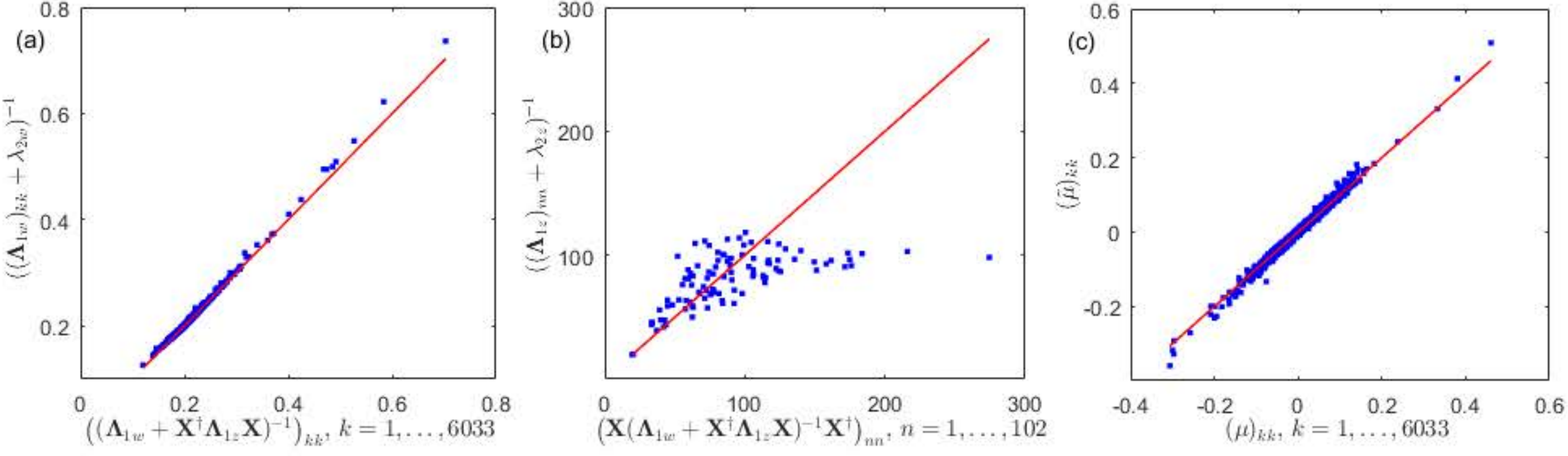}
	\vspace{-0.5cm}
	\caption{(a) Quality of approximation \eqref{approx1}. (b) Quality of approximation \eqref{approx2}. (c) Comparison of posterior means obtained by the diagonal-EP and the scalar-EP. Straight lines denote identity.}\label{fig1}\vspace{-0.1cm}
\end{figure*}

We assume that the expression level values $\{X_{nk}\}$ are irregular enough
(arising from both biological and experimental variability) to be considered ``random'' and that $N$ and $K$
are large enough to justify the application of the free probability approach. Then, Theorem~\ref{main} offers the approximations for the fixed-point equations of diagonal-EP in \eqref{fix2} as 
\begin{subequations}
	\label{approx}
	\begin{align}
	\hspace{-0.3cm}((\matr \Lambda_{1w}+\matr X^\dagger {\matr \Lambda}_{1z}\matr X)^{-1})_{kk}&\approx ((\matr\Lambda_{1w})_{kk}+\lambda_{2w})^{-1}\label{approx1} \\
	\hspace{-0.3cm}(\matr X(\matr \Lambda_{1w}+\matr X^\dagger{\matr \Lambda}_{1z}\matr X)^{-1}\matr X^\dagger)_{nn}&\approx ((\matr{\Lambda}_{1z})_{nn}+\lambda_{2z})^{-1}  \label{approx2}
	\end{align}	
\end{subequations}
where $\lambda_{2w}$ and $\lambda_{2z}$ are appropriately computed scalars which are asymptotically consistent with \eqref{solsoflambdas}. In Figure~\ref{fig1}-a) and -b) we illustrate the validity of the approximations in \eqref{approx} on the prostate cancer microarray dataset \cite{singh_gene} with $N=102$ patients and $K=6033$ genes. 
Approximation \eqref{approx} is excellent for the large matrix, but 
(as might be expected) yields cruder results for the comparably small matrix related to the small number of patients available. Nevertheless, such a mismatch  may not necessarily have a deteriorating role to the approximate posterior of $\matr w$. To show this, we compare the posterior mean using the diagonal-EP denoted by ${\matr \mu}$ and compare with the scalar-EP denoted by $\tilde{\matr \mu}$. As shown in Figure~1-c both methods yield in fact close results.\vspace{-0.1cm}
\section{Summary and Outlook}
By utilizing the concept of asymptotic freeness we have introduced a theoretical approach to EP approximate inference method. Our analysis plays a~``bridging role'' for the gap between the perspectives of machine learning and information theory communities, i.e. the diagonal- and scalar- EP approaches, respectively. We have shown a concrete application with real data (gene selection) where the free probability framework is applicable. 

For data matrices that show strong deviations from asymptotic freeness we may need to take the deviations into account by a perturbation approach. Specifically, our proof technique is based on the exact infinite sum representation of the inverse of a sum of matrices -- originally utilized by \cite{Tao12} for the additivity of R-transforms of the sum of free matrices. While freeness of matrices leads to the vanishing of most terms under the trace operation, we may keep a small, possibly increasing number of such terms to account for the deviations. This could  lead to an improvement for the scalar-EP method. 
\section*{Acknowledgment}
The authors would like to thank Daniel Hern{\'a}ndez-Lobato and Jos{\'e} M. Hern{\'a}ndez-Lobato for discussions on real data application of our approach and providing the microarray datasets and the anonymous reviewers for their suggestions and comments.
\bibliographystyle{IEEEtran}
\bibliography{IEEEabrv,mybib}  
\appendices 
\section{Proof of Theorem~1: Preliminaries}
\subsection{The R- and S- transforms}
\begin{definition}\cite{bercofree,hager3}
	Let ${\rm F}$ be a probability distribution function on the real line and $\rm G$ be its Stieltjes transform, i.e.
	\begin{equation}
	{\rm G}(z)\triangleq \int \frac{{\rm dF}(x)}{x-z}, \quad z\in \CC\backslash\RR.
	\end{equation}
 Moreover, let us denote the inverse (with respect to functional decomposition) of the Stieltjes transform by ${\rm B}(s)\triangleq {\rm G}^{-1}(s)$.  Then, the R-transform of ${\rm F}$ is given by
	\begin{equation}
	{\rm R}(s)\triangleq {\rm B}(-s)-s^{-1}. \label{Rtrs}
	\end{equation}
	Moreover, for ${\rm \tilde R}(s)\triangleq s{\rm R}(s)$ the S-transform of ${\rm F}$ is given by
	\begin{equation}
	{\rm S}(\omega)\triangleq \omega^{-1}{\rm \tilde R}^{-1}(\omega).\label{RS}
	\end{equation}
\end{definition}
 \begin{lemma}\label{pretty}
 	Let ${\rm F}$ be a probability distribution function on the real line and ${\rm R}$ be its R-transform. Moreover, let $q\triangleq \int x^{-1}{\rm dF}(x)$ and $q<\infty$. Then, we have $q{\rm R}(-q)=1$.	
 	\begin{proof}
 		We have $q=\lim_{z\to 0}{\rm G}(z)$ where	
 		${\rm G}$ denotes the Stieltjes transform of ${\rm F}$. From \eqref{Rtrs} we have
 		\begin{equation}
 		\lim_{z\to 0}{\rm R}(-{\rm G}(z))=\lim_{z\to 0}\left(z+\frac{1}{{\rm G}(z)}\right)= \frac{1}{q}.
 		\end{equation}
 		Since the R-transform is analytic in a neighborhood of zero \cite{bercofree}, we have $\lim_{z\to 0}{\rm R}(-{\rm G}(z))={\rm R}(-\lim_{z\to 0}{\rm G}(z))$. This completes the proof. 
 	\end{proof} 
 \end{lemma}
 \begin{lemma}\cite{ralf08}\label{Rinversion}
 	Let ${\rm F}_{A}$ be a probability distribution function on the real line and ${\rm R}_{A}$ be its R-transform. Furthermore, let
 	\begin{equation}
 	{\rm G}_{A^{-1}}(z)\triangleq \int \frac{{\rm dF}(x)}{x^{-1}-z},\quad z\in \CC\backslash\RR. 
 	\end{equation}
 Moreover, let ${\rm R}_{A^{-1}}(s)\triangleq{\rm G}_{A^{-1}}^{-1}(-s)-s^{-1}$. Then, we have
 	\begin{equation}
 	\frac{1}{{\rm R}_{A}(s)}= {\rm R}_{A^{-1}}\left(-{\rm R}_{A}(s)(1+s{\rm R}_{A}(s)) \right).
 	\end{equation} 
 \end{lemma}
\subsection{The Additive and Multiplicative Free Convolutions}
If a matrix $\matr A=\matr A^\dagger$ has a LED, the Stieltjes transform of the LED of $\matr A$ is defined as
\begin{equation}
{\rm G}_{\matr A}(z)\triangleq \phi((\matr A-z{\bf I})^{-1}),\quad z\in \CC\backslash\RR.\label{stiel}
\end{equation}
Also, let ${\rm B}_{\matr A}(s)\triangleq{\rm G}^{-1}_{\matr A}(s)$. Thus, ${\rm R}_{\matr A}(s)={\rm B}_{\matr A}(-s)-s^{-1}$ stands for the R-transform of the LED of $\matr A$. Similarly, ${\rm S}_{\matr A}$ stands for the S-transform of the LED of $\matr A$.

If $\matr A=\matr A^\dagger$ and $\matr B=\matr B^\dagger$ have a LED each and are asymptotically free, we have
\begin{align}
{\rm R}_{\matr A+\matr B}(s)&={\rm R}_{\matr A}(s)+{\rm R}_{\matr B}(s)\label{adf}.
\end{align}
If, in addition, $\phi(\matr A)\neq 0\neq \phi(\matr B)$ we have 
\begin{align}
{\rm S}_{\matr A\matr B}(\omega)&={\rm S}_{\matr A}(\omega){\rm S}_{\matr B}(\omega).\label{mdf}
\end{align}
The results \eqref{adf} and \eqref{mdf} are commonly referred to as the additive and multiplicative free convolutions, respectively \cite{ralfc}. 
\subsection{A local law for addition of free matrices}   
\begin{theorem}\label{LLAFM}
Let $\matr \Lambda_1$ and $\matr J=\matr J^\dagger$ be an $N\times N$ real diagonal matrix and an $N\times N$ matrix, respectively. Furthermore, let $\matr \Lambda_{2}$ be an $N\times N$ diagonal matrix with its diagonal entries introduced through
\begin{equation}
((\matr \Lambda_{1}+\matr \Lambda_{2})^{-1})_{nn}=((\matr \Lambda_1+\matr J)^{-1})_{nn}, ~~\forall n.
\end{equation}
Moreover, let $\matr Z$ be an $N\times N$ diagonal matrix and independent of $\matr \Lambda_1$ and $\matr J$. The diagonal entries of $\matr Z$ are independent and composed of $\pm 1$ with equal probabilities.

Let $\matr \Lambda_1$ and $\matr J$ have a LED each a.s. and $\matr J$ be a.s. asymptotically free of $\{\matr \Lambda_1,\matr Z\}$ as $N\to \infty$. Then, for $\chi \triangleq \phi((\matr \Lambda_1+\matr J)^{-1})$ {\new and $\lambda_2\triangleq {\rm R}_{\matr J}(-\chi)$
we have a.s. that
\begin{equation}
\{(\matr\Lambda_{2})_{11},\cdots,(\matr\Lambda_{2})_{NN}\}\overset{L^2}{\rightarrow}\lambda_2 \label{result}
\end{equation}}whenever $(\matr \Lambda_1+\lambda_2{\bf I})^{-1}$ has a.s. an asymptotically bounded spectral norm. 
\end{theorem}
\subsection*{Proof of Theorem~\ref{LLAFM}} 
For convenience we define $\matr Y\triangleq \matr \Lambda_1+\lambda_2{\bf I}$. By invoking Lemma~\ref{pretty} and the additive free convolution \eqref{adf} we have that
\begin{align}
\chi=\phi((\matr \Lambda_1+\matr J)^{-1})&= \left({\rm R}_{\matr \Lambda_1+\matr J}(-\chi) \right)^{-1}\\ 
&=\left({\rm R}_{\matr \Lambda_1}(-\chi)+\lambda_2 \right)^{-1}\\
&=\left({\rm R}_{\matr Y}(-\chi)\right)^{-1} \label{p1}\\
&=\phi(\matr Y^{-1}).\label{p2}
\end{align}
Here, $\chi<\infty$ since $\matr Y^{-1}$ is assumed to have a asymptotically bounded spectral norm (a.s.). We now define 
\begin{equation}
\matr D\triangleq (\matr \Lambda_1+\matr J)^{-1}-\matr Y^{-1}. \label{D}
\end{equation}
{\new To prove Theorem~\ref{LLAFM} it is sufficient to show (a.s.) that}
\begin{equation}
\lim_{N\to \infty}\frac{1}{N}\sum_{n=1}^{N}(D_{nn})^2=0
\end{equation}
{\new i.e. $\lim_{N\to \infty} \mathbb E[D_N^2]=0$ where $D_N$ is distributed according to the empirical distribution function of $\{D_{11},\cdots,D_{NN}\}$.} 

Note that 
${\mathbb E}[Z_{nn}Z_{kk}]=\delta_{nk}$. Hence, we have
\begin{align}
{\mathbb E}\left[{\rm Tr}((\matr Z\matr D)^2)\right]&=\frac{1}{N}\sum_{n,k=1}^N {\mathbb E}[Z_{nn}D_{nk}Z_{kk}D_{kn}]\\
&=\frac{1}{N}\sum_{i=1}^{N} (D_{nn})^2
\end{align}
Here and throughout the section expectation is taken over the random matrix $\matr Z$.  Thus, showing the condition 
\begin{equation}
\phi((\matr Z\matr D)^2)=0
\end{equation}
is sufficient for the proof. To that end we first represent the matrix $\matr D$ in a convenient mathematical formula for the algebra of asymptotic freeness, i.e. \eqref{cond}.
\begin{lemma}\label{decompos}
Let the matrices $\matr J$, $\matr Y$ and $\matr D$ be given as in \eqref{D}. Furthermore, let $\chi=\phi(\matr Y^{-1})$. Moreover, for $\matr M\in \{\matr J, \matr Y\}$ let us define
	\begin{align}
	{\matr E}_{M}\triangleq {\bf I}-\frac{1}{\chi}(\matr M-{\rm B}_{\matr M}(\chi){\bf I})^{-1}
	\end{align}
	where for short we write ${\matr E}_{M}={\matr E}_{M}(\chi)$. In particular, $\phi({\matr E}_{M})=0$. Then, for
	\begin{align}
	\matr A&\triangleq ({\bf I}-\matr E_{Y})\sum_{n=1}^{\infty}(\matr E_{J}\matr E_{Y})^{n}\label{defA}\\
	\matr B&\triangleq ({\bf I}-{\matr E}_{J})\sum_{n=1}^{\infty}(\matr E_{Y}\matr E_{J})^{n}.
	\end{align}
	we have $\matr D=\chi (\matr A+\matr B-{\matr E}_{J})$.
\end{lemma}
The proof of Lemma~\ref{decompos} is located at Appendix~\ref{prooflemma2}. 

From Lemma~\ref{decompos} we write 
\begin{align}
\frac{1}{\chi^2}\phi((\matr Z\matr D)^2)=\phi((\matr Z\matr A)^2)+\phi((\matr Z\matr B)^2)+\phi((\matr Z\matr E_J)^2)\nonumber\\
+2\phi(\matr Z\matr A\matr Z\matr B)-2\phi(\matr Z\matr A\matr Z\matr E_J)-2\phi(\matr Z\matr B\matr Z\matr E_J). \label{summand}
\end{align}
We now consider the assumption that $\matr J$ is (a.s.) asymptotically free of $\{\matr \Lambda, \matr Z\}$. This implies that $\mathcal Q_1=\{\matr E_{J}\}$ and $\mathcal Q_2\triangleq\{\matr E_Y,\matr Z\}$ are asymptotically free. Then, to show that each terms in the summand \eqref{summand} are zero, we make use of the definition of asymptotic freeness \eqref{cond} and the following specific results (in the almost sure sense)
\begin{align}
\phi(\matr E_{Y}\matr Z)&=0 ~ \text{ and }~\phi(\matr E_{Y}\matr Z\matr E_{Y})=0\label{res2}
\end{align}
where the proof of \eqref{res2} is given at Appendix~\ref{proofress}. As regards the first term in the summand \eqref{summand} it follows from \eqref{defA} that $\phi((\matr Z\matr A)^2)=0$ if for all integers $n,m\geq 1$ we have 
\begin{equation}
\phi(\matr E_{ZY}(\matr E_J\matr E_Y)^n\matr E_{ZY}(\matr E_J\matr E_Y)^m)=0 \label{simple}
\end{equation}
where for convenience we define $\matr E_{ZY}\triangleq \matr Z({\bf I}-\matr E_{Y})$. Notice that all adjacent factors in the product \eqref{simple} are polynomials belonging to the different set in the family $\{\mathcal Q_1, \mathcal Q_2\}$. Note also that $\phi(\matr E_{ZY})=0$, $\phi(\matr E_J)=0$ and $\phi(\matr E_{Y})=0$. Thus, \eqref{simple} follows directly from the definition of asymptotic freeness, i.e. from \eqref{cond}. Similarly, the other terms in the summand \eqref{summand} are zero as well. 
\subsubsection{Proof of Lemma~\ref{decompos}}\label{prooflemma2}
The proof is based on the argumentation by Tao \cite[pp. 213]{Tao12} from which he obtained an elegant derivation of the additive free convolution \eqref{adf}: Recall that ${\rm G}_{\matr M}(z)=\phi((\matr M-z{\bf I})^{-1})$. This equation can be viewed as $s=\phi((\matr M-{\rm B}_{\matr M}(s){\bf I})^{-1})$ for $s={\rm G}_{\matr M}(z)$ and thus,
\begin{equation}
(\matr M-{\rm B}_{\matr M}(s){\bf I})^{-1}=s({\bf I}-{\matr E}_{M})
\end{equation}
for some ${\matr E}_{M}={\matr E}_{M}(s)$ such that $\phi({\matr E}_{M})=0$. In particular, with a straightforward algebraic manipulation we can write \cite[pp. 213]{Tao12}
\begin{align}
(\matr J+\matr Y-({\rm B}_{\matr J}(s)+{\rm B}_{\matr Y}(s)+s^{-1}){\bf I})^{-1}~~~~\label{taorrick}\\
~~~~= s({\bf I}-\matr E_{Y})({\bf I}-\matr E_{J}\matr E_{Y})^{-1} ({\bf I}-\matr E_{J}).
\end{align} 
We are interested in this equation for $s\to\chi$: From \eqref{p1} we have ${\rm B}_{\matr Y}(\chi)=0$. Thereby, we can write \eqref{taorrick} for $s\to\chi$ as
\begin{align}
&(\matr J+\matr Y-({\rm B}_{\matr J}(\chi)+{\rm B}_{\matr Y}(\chi)+\chi^{-1}){\bf I})^{-1}\\
&=(\matr J+\matr Y-({\rm B}_{\matr J}(\chi)+\chi^{-1}){\bf I})^{-1}\\
&=(\matr J+\matr Y-\lambda_2{\bf I})^{-1}\\
&=(\matr J+\matr \Lambda_1)^{-1}. 
\end{align}
Also, ${\rm B}_{\matr Y}(\chi)=0$ leads to $\matr Y^{-1}=\chi({\bf I}-\matr E_{Y})$. Thus, we get
\begin{align}
&\matr D= (\matr J+\matr \Lambda_1)^{-1}- \matr Y^{-1}\\
&=\chi({\bf I}-\matr E_{Y})\left(({\bf I}-\matr E_{J}\matr E_{Y})^{-1} ({\bf I}-\matr E_{J})-{\bf I}\right)\\
&=\chi({\bf I}-\matr E_{Y})\left(\left({\bf I}+\sum_{n=1}^{\infty}(\matr E_{J}\matr E_{Y})^{n}\right)({\bf I}-\matr E_{J})-{\bf I}\right) \label{fin}
\end{align}
where in \eqref{fin} we expand out $({\bf I}-\matr E_{J}\matr E_{Y})^{-1}$ as \emph{formal} Neumann series. 

From \eqref{fin} we write $\matr D/\chi$ as
\begin{align}
&({\bf I}-\matr E_{Y})\left(\sum_{n=1}^{\infty}(\matr E_{J}\matr E_{Y})^{n}-\matr E_J-\matr E_J \sum_{n=1}^{\infty}(\matr E_{Y}\matr E_{J})^{n}\right)\nonumber \\
&= \matr A+({\bf I}-\matr E_{Y})\left(-\matr E_J-\matr E_J \sum_{n=1}^{\infty}(\matr E_{Y}\matr E_{J})^{n}\right)\nonumber\\
&= \matr A-\matr E_J+\left(\matr E_{Y}\matr E_{J}-({\bf I}-\matr E_{Y})\matr E_J \sum_{n=1}^{\infty}(\matr E_{Y}\matr E_{J})^{n}  \right) \nonumber \\
&=\matr A-\matr E_J+\left(\sum_{n=1}^{\infty}(\matr E_{Y}\matr E_{J})^{n}-\matr E_J\sum_{n=1}^{\infty}(\matr E_{Y}\matr E_{J})^{n} \right)\nonumber \\
&=\matr A-\matr E_J+\matr B
\end{align}
which completes the proof.
\subsubsection{Proof of \eqref{res2}}\label{proofress}
We have
\begin{align}
\mathbb{E}[{\rm Tr}(\matr E_{Y}\matr Z)]=0~\text{ and }~\mathbb{E}[{\rm Tr}(\matr E_{Y}^2\matr Z)]=0. 
\end{align}
Since $\matr Y^{-1}$ has an asymptotically bounded spectral norm so does $\matr E_{Y}$. Thus, from Kolmogorov's strong law of large numbers we have a.s. that ${\phi}(\matr E_{Y}\matr Z)=0$ and ${\phi}(\matr E_{Y}^2\matr Z)=0$.
\section{Proof of Theorem~\ref{main}}
Theorem~\ref{main}-1) follows directly from Theorem~\ref{LLAFM} such that we have
\begin{equation}
\lambda_{2w}={\rm R}_{\matr X^\dagger \matr \Lambda_{1z}\matr X}(-\chi_{w})\label{cav1}.
\end{equation}
As regards Theorem~\ref{main}-2), we first underline the following points: 
\begin{itemize}
\item By invoking the Woodbury ~matrix~inversion~lemma, for $\tilde {\matr \Sigma}\triangleq(\matr \Lambda_{1z}^{-1}+\matr X\matr \Lambda_{1w}^{-1}\matr X^\dagger)^{-1}$ we write
	\begin{align}
	(\tilde {\matr \Sigma})_{nn}&=(\matr \Lambda_{1z})_{nn}- (\matr \Lambda_{1z}^2)_{nn}(\matr X\matr \Sigma\matr X^\dagger)_{nn} \\
	&= (\matr \Lambda_{1z})_{nn}- \frac{(\matr \Lambda_{1z}^2)_{nn}}{(\matr \Lambda_{1z})_{nn}+(\matr \Lambda_{2z})_{nn}}\\
	&=\frac{1}{({\matr \Lambda_{1z}^{-1}})_{nn}+({\matr\Lambda_{2 z}^{-1}})_{nn}}\label{second}.
	\end{align}
\item A matrix $\matr A=\matr A^\dagger$ has a LED if, and only if the limiting Stieltjes transform ${\rm G}_{\matr A}(z)$ (for $z\in \CC\backslash \RR$) exists\cite{Geronimo}. On the other hand, for $\matr A$ being invertible we have \cite{ralf08}
	\begin{equation}
	{\rm G}_{\matr A^{-1}}(z)= -z^{-1}(1+z^{-1}{\rm G}_{\matr A}(z^{-1})).
	\end{equation}
Thus, $\matr \Lambda_{1w}^{-1}$ and $\matr \Lambda_{1z}^{-1}$ have LEDs as $\matr \Lambda_{1w}$~and~$\matr \Lambda_{1z}$ do. 
\item We write $\matr\Lambda_{1z}^{-1}$ of the form $\sum_{n=0}^{N-1}\alpha_n \matr \Lambda_{1z}^n$ for some $\alpha_n$ (depending on the spectral moments ${\rm Tr}(\matr \Lambda_{1z}^k)$ up to some order). Thereby, it follows from Definition~\ref{deffree} that $\matr X\matr \Lambda_{1w}^{-1}\matr X^\dagger$ and $\matr\Lambda_{1z}^{-1}$ are asymptotically free as the former matrix is assumed to be free of the inverse of the latter. 
\end{itemize}
From these remarks Theorem~\ref{main}-2) follows from Theorem~\ref{LLAFM} such that we have
\begin{align}
\lambda_{2z}=\frac{1}{{\rm R}_{\matr X\matr \Lambda_{1w}^{-1}\matr X^\dagger}(-\tilde\chi_{z})} ~\text{ with } \tilde \chi_{z}\triangleq\phi(\tilde {\matr \Sigma})\label{cav2}.
\end{align}
Here, we note that since $(\matr\Lambda_{1z}+\lambda_{2z}{\bf I})^{-1}$ has an (a.s.) asymptotically bounded spectral norm  so does
\begin{equation}
(\matr\Lambda_{1z}^{-1}+\lambda_{2z}^{-1}{\bf I})^{-1}=\lambda_{2z}{\bf I}-\lambda_{2z}^2(\matr\Lambda_{1z}+\lambda_{2z}{\bf I})^{-1} \label{resecond}.
\end{equation}
Moreover, from \eqref{resecond} it follows that
\begin{equation}
\tilde \chi_{\rm z}=\lambda_{2z}(1-\lambda_{2z}\chi_{z}). \label{relzz}
\end{equation}
To complete the proof we only need to show that $\lambda_{2w}$ in \eqref{cav1} and $\lambda_{2z}$ in \eqref{cav2} fulfill the system of equations in \eqref{solsoflambdas}. To this end we outline two remarks whose proofs are given at the end of the appendix, for convenience. 

\begin{remark}
Let $\matr \Lambda_{1w}$, $\matr \Lambda_{1z}$ and $\matr X^\dagger \matr X$ have a LED each as $N,K\to \infty$ with $N/K=O(1)$. Furthermore, let the pairs
$\matr X\matr X^\dagger$~\&~$\matr \Lambda_{1z}$ and $\matr X^\dagger \matr \Lambda_{1z}\matr X$~\&~$\matr \Lambda_{1w}$ form an asymptotically free family each.  Moreover, let $\chi_{w}\triangleq\phi((\matr \Lambda_{1w}+\matr X^\dagger \matr \Lambda_{1z}\matr X)^{-1})$ be finite and $\lambda_{2w}$ be as in \eqref{cav1}. Then, we have 
	\begin{equation}
	\chi_{w}=\frac{1}{\lambda_{1w}+\lambda_{2 w}}=\phi((\lambda_{1w}{\bf I}+\lambda_{1z}\matr X^\dagger \matr X)^{-1})
	\end{equation}
	where for $\alpha\triangleq N/K$ we define
	\begin{equation}
	\lambda_{1w}\triangleq{\rm R}_{\matr \Lambda_{1 w}}(-\chi_{w}) ~\text{ and }~\lambda_{1 z}\triangleq\left({\rm S}_{\matr \Lambda_{1 z}}\left(-\frac{\lambda_{2w}\chi_{ w}}{\alpha}\right)\right)^{-1}.\label{lambdas}
	\end{equation} 	
\end{remark}
\begin{remark}
Let $\matr \Lambda_{1w}$, $\matr \Lambda_{1z}$ and $\matr X^\dagger \matr X$ have a LED each as $N,K\to \infty$ with $N/K=O(1)$. Furthermore, let the pairs $\matr X^\dagger\matr X$~\&~$\matr \Lambda_{1w}$ and $\matr X\matr \Lambda_{1w}^{-1}\matr X^\dagger$~\&~$\matr \Lambda_{1z}$ form an asymptotically free family each. Moreover, let $\tilde \chi_{z}\triangleq\phi(({\matr \Lambda_{1z}^{-1}}+\matr X{\matr \Lambda_{1w }^{-1}}\matr X^\dagger)^{-1})$ be finite and $\chi_z\triangleq (1-\tilde \chi_z/\lambda_{2z})/\lambda_{2z}$ (i.e. $\tilde \chi_z$ is of the form \eqref{relzz}) where $\lambda_{2z}$ is as in \eqref{cav2}. 	Then, we have 
\begin{equation}
\chi_{z}=\frac{1}{\tilde\lambda_{1z}+\lambda_{2 z}}=\phi(\matr X(\tilde\lambda_{1w}{\bf I}+\tilde\lambda_{1z}\matr X^\dagger \matr X)^{-1}\matr X^\dagger)\label{remf}
\end{equation}
where for $\alpha\triangleq N/K$ we define
\begin{equation}
{\tilde{\lambda}_{1z}\triangleq{\rm R}_{\matr \Lambda_{1 z}}(-\chi_{z})~\text{ and }~\tilde\lambda_{1w}\triangleq{\rm S}_{\matr \Lambda_{1w}^{-1}}\left(-\frac{\alpha\tilde\chi_{ z}}{\lambda_{2z}}\right)}.\label{lambdass}
\end{equation}
\end{remark}
We thereby complete the proof by showing that the two solutions (i) $\tilde \lambda_{1w}=\lambda_{1w}$ and (ii) $\tilde\lambda_{1 z}=\lambda_{1z}$ are \emph{consistent}. To this end, we postulate that
\begin{equation}
\lambda_{2w}\chi_{w}{=}\alpha(1-\lambda_{2z}\chi_{z}).\label{conj}
\end{equation}
We first obtain (i) and (ii) under the hypothesis \eqref{conj}; then we show the hypothesis holds when (i) and (ii) hold. 

By using respectively the first equality of \eqref{lambdas}, \eqref{conj} and the first equalities in \eqref{remf} and \eqref{lambdass} we obtain that
\begin{align}
\frac{1}{\lambda_{1z}}&={\rm S}_{\matr \Lambda_{1 z}}(-(1-\lambda_{2 z}\chi_{z}))\\
&={\rm S}_{\matr \Lambda_{1z}}(-\tilde\lambda_{1 z}\chi_{ z})\\
&={\rm S}_{\matr \Lambda_{1z}}(-{\rm R}_{\matr \Lambda_{1 z}}(-\chi_{z})\chi_{z}).
\end{align}
From \eqref{RS}, this implies that $\lambda_{1z}={\rm R}_{\matr \Lambda_{1z}}(-\chi_{z})$, i.e. $\lambda_{1 z}=\tilde\lambda_{1z}$. 

By using respectively the second equality of \eqref{lambdas}, \eqref{relzz}, \eqref{conj} and \eqref{RS} we obtain that
\begin{align}
\tilde \lambda_{1w}&={\rm S}_{\matr \Lambda_{1 w}^{-1}}\left(-\alpha (1-\lambda_{2z}\chi_{z})\right)\\
&= {\rm S}_{\matr \Lambda_{1w}^{-1}}\left(-\lambda_{2 w}\chi_{w}\right)\\
&={\rm S}_{\matr \Lambda_{1w}^{-1}}\left(-(1-\lambda_{1 w}\chi_{w})\right)\\
&=\frac{1}{{\rm R}_{\matr \Lambda_{1w}^{-1}}(\tilde\lambda_{1w}(1-\lambda_{1 w}\chi_{w}))}. \label{alfin}
\end{align}
From Lemma~\ref{Rinversion} the equality $\tilde\lambda_{1w}={\rm R}_{\matr \Lambda_{1w}}(-\chi_{w})$, i.e. $\tilde\lambda_{1w}=\lambda_{1w}$, fulfills \eqref{alfin} and thereby the solution (ii) is consistent too. 

We complete the proof by showing  that the hypothesis \eqref{conj} holds when (i) and (ii) hold:
\begin{align}
\hspace{-0.13cm}\lambda_{2w}\chi_{w}&=1-\lambda_{1w}\chi_{w} \\
&=\phi({\bf I}-\lambda_{1w}(\lambda_{1w}{\bf I}+\lambda_{1z}\matr X^\dagger \matr X)^{-1}) \\
&=\alpha\phi({\bf I}-\lambda_{1w}(\lambda_{1w}{\bf I}+\lambda_{1z}\matr X\matr X^\dagger)^{-1}) \\
&=\alpha\phi({\bf I} -\lambda_{1z}^{-1}(\lambda_{1z}^{\small -1}{\bf I}+ {\lambda_{1w}^{\small -1}}\matr X \matr X^\dagger)^{-1})\\
&=\alpha\phi({\bf I}-({\bf I}-\lambda_{1z}\matr X(\lambda_{1w}+\lambda_{1z}\matr X^\dagger \matr X )^{-1}\matr X^\dagger))\label{pfinal}\\
&=\alpha(1 -(1-\lambda_{1z}\chi_{z}))\\
&=\alpha(1-\lambda_{2z}\chi_{z}).
\end{align} 
Here \eqref{pfinal} follows from the Woodbury~matrix~inversion~lemma. 
\subsection{Proof of Remark~1}\label{rem1p}
From Lemma~\ref{pretty} we have
\begin{equation}
\chi_{w}=(\lambda_{1w}+\lambda_{2w})^{-1}.
\end{equation}
With the scaling property of the R-transform (see \cite[Table~4]{ralfc}), the equality ${\lambda}_{2w}={\rm R}_{\matr X^\dagger \matr \Lambda_{1z}\matr X}(-\chi_{w})$ holds if, and only if
\begin{equation}
1={\rm R}_{\frac{1}{{\lambda}_{2w}}\matr X^\dagger \matr \Lambda_{1z}\matr X}(-{\lambda}_{2w} \chi_{w}).
\end{equation}
We express this result in terms of the S-transform via \eqref{RS} as:
\begin{align}
1&= {\rm S}_{\frac{1}{{\lambda}_{2w}}\matr X^\dagger \matr \Lambda_{1z}\matr X}(-{\lambda}_{2w}\chi_{w})\\
&={\rm S}_{\frac{1}{{\lambda}_{2w}}\matr \Lambda_{1z}\matr X\matr X^\dagger}\left(-\frac{{\lambda}_{2w}\chi_{w}}{\alpha}\right)\\
&=\underbrace{{\rm S}_{\matr \Lambda_{1z}}\left(-\frac{{\lambda}_{2w}\chi_{w}}{\alpha}\right)}_{\lambda_{1z}^{-1}}{\rm S}_{\frac{1}{{\lambda}_{2w}} \matr X\matr X^\dagger}\left(-\frac{{\lambda}_{2w}\chi_{w}}{\alpha}\right)\label{rmfc}\\
&={\rm S}_{\frac{\lambda_{1z}}{{\lambda}_{2w}} \matr X\matr X^\dagger}\left(-\frac{{\lambda}_{2w}\chi_{w}}{\alpha}\right)\\
&={\rm S}_{\frac{\lambda_{1z}}{{\lambda}_{2w}}\matr X^\dagger \matr X}(-{\lambda}_{2w}\chi_{w})\label{Sscale}.
\end{align}
Here \eqref{rmfc} follows from (c) and the multiplicative free convolution \eqref{mdf}. We now reuse \eqref{RS} to express \eqref{Sscale} in terms of the R-transform as
\begin{equation}
{\lambda}_{2w}={\rm R}_{\lambda_{1z}\matr X^\dagger\matr X}(-\chi_{w}).
\end{equation}
This completes the proof. 
\subsection{Proof of Remark~2}\label{rem2p}
We make use of Remark~1 for a different ``system'' as
\begin{align}
\phi((\matr \Lambda_{1z}^{-1}+\matr X\matr \Lambda_{1w}^{-1}\matr X^\dagger)^{-1})=\phi((\tilde \lambda_{1z}^{-1}{\bf I}+ {\tilde\lambda_{1w}^{-1}}\matr X \matr X^\dagger)^{-1}) \label{ff} 
\end{align}
such that we have 
\begin{equation}
\tilde\chi_{z}=\frac{1}{\tilde\lambda_{1z}^{-1}+\lambda_{2 z}^{-1}}=\phi((\tilde\lambda_{1 z}^{-1}{\bf I}+\tilde\lambda_{1w}^{-1}\matr X\matr X^\dagger)^{-1}) \label{rems}
\end{equation}
where
\begin{equation}
\tilde\lambda_{1z}\triangleq\frac{1}{{\rm R}_{\matr \Lambda_{1z}^{-1}}(-\tilde \chi_z)}~\text{~and~} \tilde\lambda_{1w}\triangleq{\rm S}_{\matr \Lambda_{1w}^{-1}}\left(-\frac{\alpha\tilde\chi_{ z}}{\lambda_{2z}}\right)\label{r2res}.
\end{equation}
By invoking the first equality of \eqref{rems} to the definition of $\chi_z$ we get $\chi_{z}=1/(\tilde\lambda_{1z}+\lambda_{2z})$ and thereby we have
\begin{align}
\tilde\chi_{z}&=\lambda_{2z}(1-\lambda_{2z}\chi_{z}) \\
 &=\tilde\lambda_{1z}(1-\tilde \lambda_{1z}\chi_{z}). \label{tildchiz}
\end{align}
Then, by the Woodbury matrix inversion lemma, \eqref{rems} implies \eqref{remf}. Furthermore,  plugging \eqref{tildchiz} into the first equality of \eqref{r2res} we have
\begin{equation}
	\frac{1}{\tilde \lambda_{1z}}= {\rm R}_{\matr \Lambda_{1z}^{-1}}(-\tilde\lambda_{1z}(1-\tilde \lambda_{1 z}\chi_{z})).
\end{equation}
From Lemma~\ref{Rinversion} this implies that $\tilde \lambda_{1z}= {\rm R}_{\matr \Lambda_{1z}}(-\chi_{z})$. This completes the proof. 

\end{document}